%% file: main.tex
\documentclass[ 10pt,conference]{IEEEtran}
\IEEEoverridecommandlockouts
\usepackage{cite}
\usepackage{textcomp}
\usepackage{xcolor}
\usepackage{stfloats}
\usepackage{algorithmic}
\usepackage{graphicx}
\usepackage{textcomp}
\usepackage{xcolor}
\usepackage{multirow}
\usepackage{array}
\usepackage{colortbl}
\usepackage{verbatim}
\usepackage{url}
\usepackage{float}
\usepackage{tcolorbox}
\usepackage[linesnumbered,lined,boxed,commentsnumbered]{algorithm2e}
\usepackage{booktabs}
\usepackage{amsmath, bm}
\usepackage{listings}
\usepackage{subfigure}
\usepackage{makecell}
\usepackage{multirow}
 \usepackage{fancyhdr}

\newcommand{\tool}{\texttt{DeepPseudo}}

\begin{document}


\title{Fine-grained Pseudo-code Generation Method via  Code Feature Extraction and Transformer}

\author{\IEEEauthorblockN{Guang Yang\IEEEauthorrefmark{2},  
Yanlin Zhou\IEEEauthorrefmark{2}, 
Xiang Chen\IEEEauthorrefmark{2}\IEEEauthorrefmark{3}\IEEEauthorrefmark{4}\IEEEauthorrefmark{1},  Chi Yu\IEEEauthorrefmark{2}}
\IEEEauthorblockA{\IEEEauthorrefmark{2}\textit{School of Information Science and Technology},
\textit{Nantong University}, China\\
\IEEEauthorrefmark{3}\textit{Key Laboratory of Safety-Critical Software (Nanjing University of Aeronautics and Astronautics)},\\
\textit{Ministry of Industry and Information Technology}, China\\
\IEEEauthorrefmark{4}\textit{The Key Laboratory of Cognitive Computing and Intelligent Information Processing of Fujian Education Institutions},\\
\textit{Wuyi University}, China\\
Email: 1930320014@stmail.ntu.edu.cn, 1159615215@qq.com, xchencs@ntu.edu.cn, yc\_struggle@163.com}
}


\maketitle

\begingroup
\renewcommand{\thefootnote}{}
\footnotetext[1]{\IEEEauthorrefmark{1} Xiang Chen is the corresponding author.}
\endgroup

\begin{abstract}
Pseudo-code written by natural language is helpful for novice developers' program comprehension. However, writing such pseudo-code is time-consuming and laborious. Motivated by the research advancements of sequence-to-sequence learning and code semantic learning, we propose a novel deep pseudo-code generation method {\tool} via code feature extraction and Transformer. In particular, {\tool} utilizes a Transformer encoder to perform encoding for source code and then use a code feature extractor to learn the knowledge of local features. Finally, it uses a pseudo-code generator to perform decoding, which can generate the corresponding pseudo-code. 
We choose two corpora  (i.e., Django and SPoC) from real-world large-scale projects as our empirical subjects. We first compare {\tool} with seven state-of-the-art baselines from pseudo-code generation and neural machine translation domains in terms of four performance measures. Results show the competitiveness of {\tool}. Moreover, we also analyze the rationality of the component settings in {\tool}. 

\end{abstract}


\input{1introduction}
\input{2background}

\input{3approach}

\input{4setup}

\input{5result}
\input{6discuss}
\input{7threats}
\input{8conclusion}

\section*{Acknowledgment}

Guang Yang and Yanlin Zhou have contributed equally to this work and they are co-first authors.
This work is supported in part by National Natural Science Foundation of
China (Grant nos. 61872263 and  61202006), The Open Project of Key Laboratory of Safety-Critical Software for Nanjing University of Aeronautics and Astronautics, Ministry of Industry and Information Technology (Grant No. NJ2020022), the Open Project Program of The Key Laboratory of Cognitive Computing and Intelligent Information Processing of Fujian Education Institutions (KLCCIIP201802).
.




\bibliographystyle{IEEEtran}
\bibliography{mylib}
\end{document}

%% file: 1introduction.tex
\section{Introduction}
\label{sec:intro}

\begin{figure*}[htbp]
	\centering
    \vspace{-1mm}
	\includegraphics[width=0.95\textwidth]{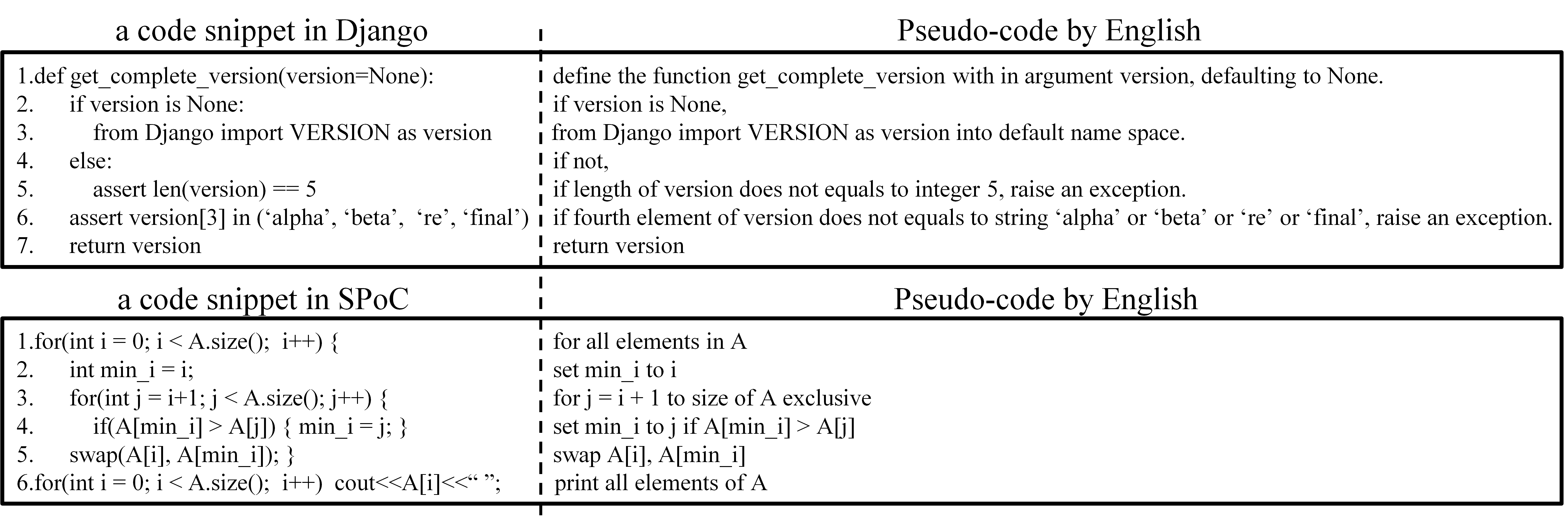}
	\caption{The example of the code snippet and its corresponding pseudo-code written in English from two corpora  Django and SPoC}
    \vspace{-1mm}
	\label{fig:snippet}
\end{figure*}

Comments written by natural language can help novice developers' program comprehension since these developers may not be familiar with the grammar of the programming languages and the domain knowledge of the projects. 
In this study, we solve the problem of line-to-line pseudo-code generation by building a neural network model. 
In the algorithms' textbook, pseudo-code is often used to introduce the problem's solution since the pseudo-code has higher readability than the source code. Then the developers can write the source code quickly according to the description of the pseudo-code. 
Moreover, for program comprehension, the developers can easily understand the pseudo-code than the code. Finally, novice developers can also learn how to write code according to the description of the pseudo-code. 
Fig.~\ref{fig:snippet} shows the example of the  code snippet (in the left part of Fig.~\ref{fig:snippet}) and its corresponding pseudo-code in English (in the right part of Fig.~\ref{fig:snippet}) in our used two parallel corpora Django and SPoC.

However, most of the project's source code has no corresponding pseudo-code since writing pseudo-code is time-consuming and laborious. If pseudo-code can be generated automatically, the developers can substantially reduce the corresponding efforts.

Motivated by the research progress of neural machine translation (NMT) and code semantic learning, we propose a deep pseudo-code generation method {\tool} via code feature extraction and Transformer~\cite{vaswani2017attention}.
In particular, {\tool} utilizes a Transformer encoder to perform encoding for source code and uses a code feature extractor to
learn the knowledge of local features. Then it uses a pseudo-code generator to perform decoding, which can generate the corresponding pseudo-code. 

The main characteristic of our proposed method {\tool} is the usage of code feature extractor via the convolutional neural network (CNN). 
Inspired by the idea of multi-model fusion (i.e., different models with different structures often learn different features)~\cite{ding2020fusion}, we fuse the local feature representation learned by the code feature extractor with the global context representation learned by the Transformer encoder to improve the generalization ability of the model. Moreover, to ensure the robustness of the model, we add adversarial training data to the embedding layers of the encoder and decoder by using the method PGD (Projected Gradient Descent)~\cite{2017Towards}. The inclusion of the adversarial training data can both improve the robustness of the model in response to malicious adversarial samples and act as a regularization to reduce the overfitting problem~\cite{2019FreeLB}.
Finally, we also consider different attention mechanisms to further improve the performance of {\tool}.

To verify the effectiveness of our proposed method {\tool}, we choose two corpora (i.e., Django~\cite{oda2015learning} and SPoC~\cite{kulal2019spoc}) gathered from real-world large-scale projects as our experimental subjects. 
We first compare {\tool} with seven state-of-the-art baselines. Three baselines are chosen from the pseudo-code generation domain, and the remaining baselines are chosen from neural machine translation domains. 
We conduct performance comparison in terms of four performance measures, which have been used in previous studies on code comment generation~\cite{hu2020deep}\cite{zhang2020retrieval}.
Results show the competitiveness of our proposed method.
Then, we also design the research questions to verify the component setting rationality in {\tool}, such as the usage of code feature extractor and the chosen attention mechanism.
The final results show that the usage of the code feature extractor and the norm-attention mechanism can achieve the best performance in {\tool}.

To our best knowledge, we summarize the main contributions as follows.

\begin{itemize}
  \item We propose a method {\tool} based on code feature extraction and Transformer. In the encoder part, {\tool} utilizes both Transformer encoder and code feature extractor to perform encoding for source code. In the decoder part, {\tool} resorts to the beam search algorithm for pseudo-code generation.
  
  \item We choose two corpora from real-world project as our experimental subject. Comparison results with seven baselines show the competitiveness of {\tool} in terms of four performance measures. 

  \item We share our scripts, trained model, and  corpora\footnote{\url{https://github.com/NTDXYG/DeepPseudo}} to facilitate the replication of our study and encourage more follow-up studies on this research topic.
\end{itemize}

The rest of this paper is organized as follows.
Section~\ref{sec:related} summarized the related work for automated pseudo-code generation and code comment generation.
Section~\ref{sec:method} shows the framework of our proposed method {\tool} and its details for each component.
Section~\ref{sec:setup} illustrates the experimental setup, including research questions, experimental subjects, performance measures, implementation details, and running platform.
Section~\ref{sec:results} performs result analysis for each research question.
Section~\ref{sec:discuss} investigates the influence of different parameter values on {\tool}.
Section~\ref{sec:threat} analyzes potential threats to the validity of our empirical studies.
Section~\ref{sec:conclusion} concludes our study and provides potential future directions.

%% file: 2background.tex
\section{Related Work}
\label{sec:related}

In this section, we first summarize the related work for code comment generation.
Then we summarize the related work for pseudo-code generation.
Finally, we emphasize the novelty of our study.

\subsection{Code Comment Generation}

Similar to pseudo-code generation, code comment generation also aims to generate comments for the target code snippet. The comment aims to summarize the  code snippet and describe the code's functionality and purpose. 
In the early stage, researchers mainly focused on the template-based methods and information retrieval-based methods. In recent years, researchers mainly considered deep learning-based methods.

Iyer et al.~\cite{iyer2016summarizing} first proposed a deep learning-based method code-NN.
Allamanis et al.~\cite{allamanis2016convolutional} considered a convolutional attention network.
Zheng et al.~\cite{zheng2017code} introduced a code attention mechanism.
Liang and Zhu~\cite{liang2018automatic} considered Code-RNN and Code-GRU for encoder and decoder.
Hu et al.~\cite{hu2018deep}\cite{hu2020deep} proposed the method DeepCom by analyzing abstract syntax trees (ASTs).
Leclair et al.~\cite{leclair2019neural} proposed the method ast-attendgru, which combines words from code with the code structure from ASTs.
Leclair et al.~\cite{leclair2020improved} then used a graph neural network (GNN), which can effectively analyze the AST structure.
Jiang et al.~\cite{jiang2017automatically} automatically generated commit messages via diffs by using a neural machine translation algorithm.
Then, Xu et al.~\cite{xu2019commit} utilized the copying mechanism, and Liu et al.~\cite{liu2020atom} utilized information retrieval-based to further improve the message quality. Liu et al.~\cite{liu2019automatic} aimed to generate pull request descriptions.
Other researchers aim to utilize different methods to improve the model performance.
Wan et al.~\cite{wan2018improving}  proposed the method Hybrid-DRL, which considers hybrid code representation and deep reinforcement learning.
Then, Wang et al.~\cite{wang2020reinforcement} further extended the method Hybrid-DRL.
Hu et al.~\cite{hu2018summarizing} proposed the method TL-CodeSum, which can utilize API knowledge learned in a related task to improve the quality of code comments.
Zhang et al.~\cite{zhang2020retrieval} proposed a retrieval-based neural code comment generation method. This method enhances the model with the most similar code segments retrieved from the training set from syntax and semantics aspects.

\subsection{Pseudo-code Generation}

Code comment generation aims to generate the description of function and purpose for the code and reduces developers' time for reading code.
While the pseudo-code generation can help novice developers understand the source code's details at the fine-grained level. According to pseudo-code, the developers can understand what each statement does.


To generate pseudo-code, Oda et al.~\cite{oda2015learning,fudaba2015pseudogen} resorted to statistical machine translation.
They mainly considered phrase-based machine translation and tree-to-string machine translation frameworks.
Hajie et al.~\cite{haije2016automatic} implemented a sequence-to-sequence model via LSTM to generate comments for code fragments.
Alhefdhi et al.~\cite{alhefdhi2018generating} proposed an approach to automatically generate pseudo-code from source code by using LSTM and Neural Machine Translation. 
Xu et al.~\cite{xu2018automatic} treated the pseudo-code generation task as a language translation task and considered a sophisticated neural machine translation model and attention seq2seq model for this task.
Li et al.~\cite{dong2018coarse} considered the replication mechanism and sketch generation for fine-grained semantic learning of code.
Rai et al.~\cite{rai2019generation} proposed an approach to generate pseudo-code from the python source code. In the first step, they converted python code into XML code for better code information extraction. Next, important information extracted from the XML code was later used to generate  pseudo-code with the help of pseudo-code templates.
Xiong et al.~\cite{xiong2020code2text} proposed the method Code2Text. The pseudo-code is generated by constructing two encoders to learn the syntactic and semantic information of the code respectively.
Deng et al.~\cite{deng2020code} proposed a type-aware sketch-based seq2seq learning method. This method can generate the natural language description of the source code, which combines the type of the source code and the mask mechanism with the LSTM.

\subsection{Novelty of Our Study}
In this study, we want to resort to the deep learning-based method to generate pseudo-code. 
Compared with previous studies,  {\tool} uses the Transformer’s encoder to
learn global code semantics and adds a stacked convolutional network (i.e., code feature extractor) to extract local code features. 
Finally, in the decoding layer, {\tool} used the Beam search algorithm
to generate fine-grained pseudo-code for each line of code based on learned code semantic.

%% file: 3approach.tex
\section{Our Proposed Method {\tool}}
\label{sec:method}

\subsection{Framework of Our Proposed Method}

\begin{figure*}[htbp]
	\centering
  \vspace{-1mm}
	\includegraphics[width=1\textwidth]{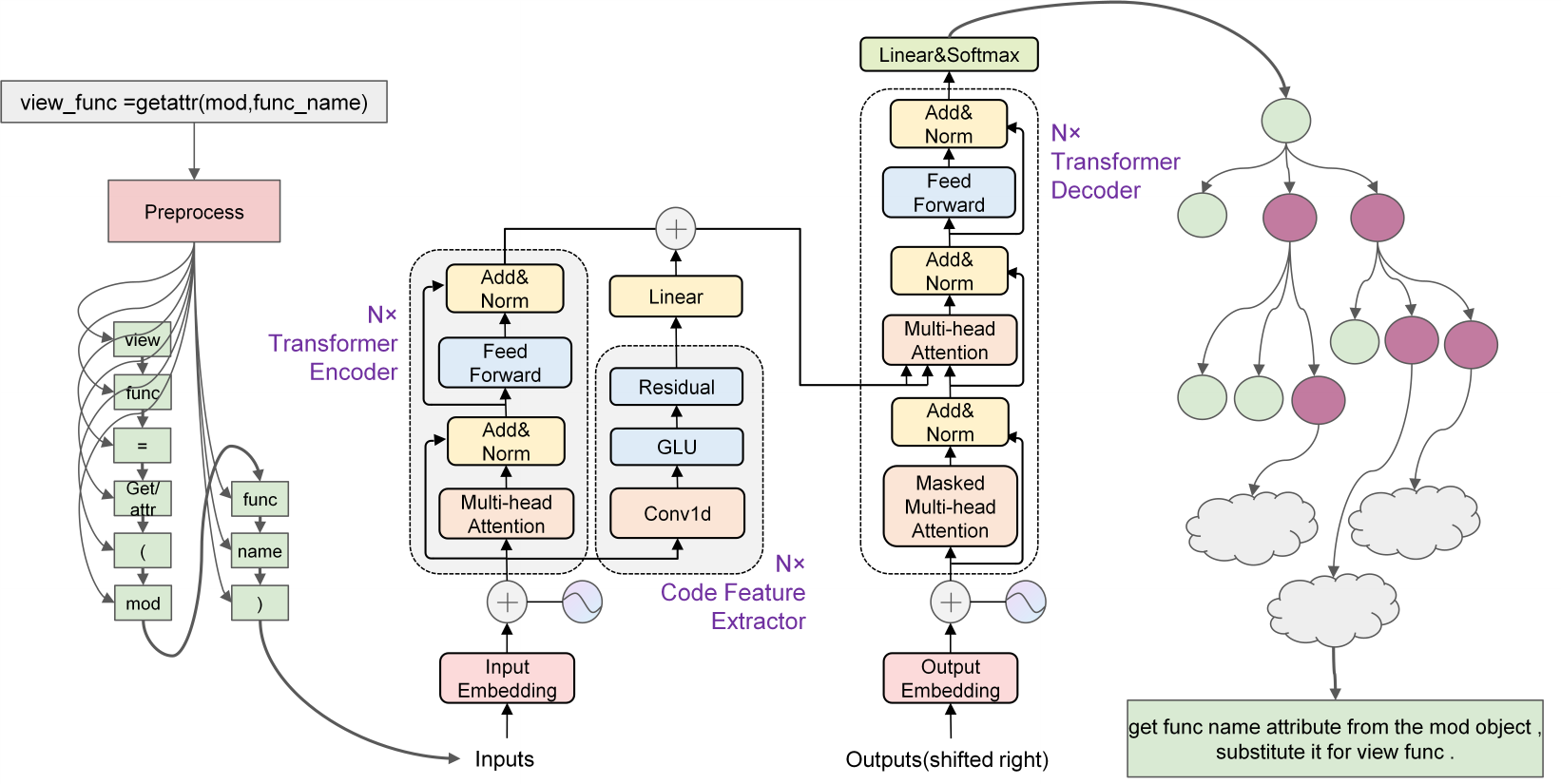}
	\caption{Framework of our proposed method {\tool}}
  \vspace{-1mm}
	\label{fig:framework}
\end{figure*}

The overall framework of {\tool} can be found in Fig.~\ref{fig:framework}.
In this figure, we can find
{\tool} includes three components:
transformer encoder, code feature extractor, and pseudo-code generator.
Then we can use the gathered corpora to train the model via this framework.
Specially, we use the method PGD (Projected Gradient Descent)~\cite{2017Towards} to perturb the word embedding layer of the encoder and decoder to improve the generalization ability of pseudo-code generation. 
Finally, we use an example in Fig.~\ref{fig:framework} to show the prediction process. 
Here, the input of the source code is \textbf{``view\_func =getattr(mod, func\_name)
"}, and the generated pseudo-code is \textbf{``get func name attribute from the mod object, substitute it for view func."}.


\subsection{Transformer Encoder}

Let $X=(x_{1},x_{2},\cdots,x_{n})$ be the input (i.e., the tokens of the code) of Transformer encoder~~\cite{vaswani2017attention}\cite{cao2021automated}. Through word embedding layer and positional encoding, {\tool} can get a vector $X$:
\begin{equation}
X = Embedding(X) + PositionalEncoding(X)
\end{equation}

Then, {\tool} inputs the vector $X$ into the multi-head attention layer.
The attention function is computed simultaneously based on three matrices: $Q$ (queries), $K$ (keys), and $V$ (values).

\begin{equation}
Q, K, V=Linear(X)
\end{equation}

\begin{equation}
X_{attention}=Attention(Q, K, V)
\end{equation}

Later, {\tool} uses residual connection~\cite{he2016deep} and layer normalization~\cite{ba2016layer} to make the matrix operation dimension consistent and normalize the hidden layer in the network to a standard normal distribution, which can accelerate the model training speed and the convergence.

\begin{equation}
X_{attention}=X + X_{attention}
\end{equation}

\begin{equation}
X_{attention}=LayerNorm(X_{attention})
\end{equation}

In the fourth step, {\tool} passes the feed-forward layer and two linear mapping layers. Then {\tool} uses the activation function to generate the vector $X_{hidden}$.

\begin{equation}
X_{hidden}=Activate(Linear(Linear(X_{attention})))
\end{equation}

Finally, {\tool} performs a residual connection~\cite{he2016deep} and layer normalization~\cite{ba2016layer} to obtain the final context vector 
$C=X_{hidden}=(c_{1},c_{2},\cdots,c_{n})$.

\begin{equation}
X_{hidden}=X_{hidden} + X_{attention}
\end{equation}

\begin{equation}
C=X_{hidden}=LayerNorm(X_{hidden})
\end{equation}

\subsection{Code Feature Extractor}

For the pseudo-code generation problem, we additionally introduced the code feature extractor (CFER) to effectively capture the source code's local feature information as external knowledge for knowledge fusion with Transformer, which can not be captured by the Transformer Encoder. Therefore, CFER is attached to the original Transformer Encoder in our proposed method {\tool}. 

Recently, more studies~\cite{hu2018deep,hu2018summarizing}  on code comment generation resorted to fusing semantic and syntactic information learned from the code to improve the quality of the generated code comments. 
In particular, they all aimed to extract the information from the abstract syntax tree (AST) of code fragments for syntactic representation learning. However, extracting this kind of information may be not feasible in pseudo-code generation tasks since the code line is fine-grained. For example, the code statement ``from logging import NullHandler", which imports a third-party library, cannot extract its AST information. Therefore, we aim to combine CNN with Transformer  in the field of pseudo-code generation.

First, we utilized the local feature information extracted by CNN via static convolution kernel. According to previous studies~\cite{cordonnier2019relationship}\cite{gong-etal-2018-convolutional}, CNN and self-attention differ in the extraction of feature information. The feature vectors of different tokens in a Conv Layer have only additive and subtractive operations on each other. Feature vectors of different tokens in self-attention have inner product operations on each other. The focus of self-attention is more on how to decide where to project more attention in a global context, whereas convolution in CNNs focuses more on obtaining another local representation of the input text. Then, We refer to the use of convolution in the ConvS2S model for local feature extraction by using cascading one-dimensional convolution to obtain local feature knowledge of the source code line. We fuse the feature representation learned by the local code feature extractor with the global context feature learned by the Transformer encoder to improve the performance of the model.

In particular, we utilize the CNN-based structure proposed by Gehrin et al.~\cite{gehring2017convolutional}. The stack CNN can form a hierarchical representation, and every convolutional layer includes a one-dimensional convolution, a gated linear unit (GLU)~\cite{2016Language}, and a residual connection.

\begin{equation}
F=X + GLU(Conv(X))
\end{equation}

The input of CFER (i.e., the vector $X$ obtained through the word embedding layer and positional encoding) is shared with the Transformer encoder. The next step of CFER is to enter the convolutional layer for feature extraction. After completing the extraction, a linear layer is used to make the feature vector $F$, which has the same dimension as the context vector $C$. Finally, the two features are merged through matrix addition, and the code representation vector $Z$ is obtained.

\begin{equation}
Z = C + Linear(F)
\end{equation}

\subsection{Pseudo-code Generator}
Previous studies~\cite{stahlberg2019nmt} showed generating the text by using the neural network's maximum probability distribution often leads to low-quality results. Recently, most studies~\cite{hu2018deep}\cite{hu2018summarizing} resorted to beam search, and these studies can achieve state-of-the-art performance on text generation tasks. Therefore we also use the Transformer Decoder and  \textbf{Beam Search} algorithm~\cite{wiseman2016sequence} to generate pseudo-code for the source code in our proposed method {\tool}. The deep learning model has poor interpretability, but the effectiveness of the beam search algorithm has been verified via theory analysis, and the beam search algorithm has an obvious connection to a  cognitive science theory (i.e., the uniform information density hypothesis)~\cite{meister2020if}.

%% file: 4setup.tex
\section{Experimental Setup}
\label{sec:setup}

In our empirical studies, we aim to investigate the following three research questions (RQs).

\noindent\textbf{RQ1: Can our proposed method {\tool} outperform state-of-the-art baselines?}

\noindent\textbf{Motivation.}
In this RQ, we first want to show the effectiveness of our proposed method {\tool} by comparing state-of-the-art baselines on pseudo-code generation. Here we mainly consider 
statistical machine translation-based methods proposed by Oda et al.~\cite{oda2015learning} and a type-aware sketch-based sequence-to-sequence method proposed by Deng et al.~\cite{deng2020code}. Moreover, we also want to compare our proposed method {\tool} with other state-of-the-art sequence-to-sequence methods (i.e., RNN based methods with/without the attention mechanism, ConvS2S, and Transformer).

\noindent\textbf{RQ2: Whether using the code feature extractor component can improve the performance of {\tool}?}

\noindent\textbf{Motivation.} After analyzing the pseudo-code generation problem, we additional add the code feature extractor component in our proposed method {\tool}. Therefore, we want to verify whether using the CFER component to extract code feature information can improve the performance of {\tool} after comparing the method {\tool} without the CFER component.

\noindent\textbf{RQ3: What is the effect of the attention mechanism on our proposed method {\tool}?}

\noindent\textbf{Motivation.}
The Transformer is based on a multi-head attention mechanism.
Each input is divided into multiple heads, and each head uses the attention mechanism.
Therefore, we want to investigate the impact of the different attention mechanisms on our proposed method {\tool}. In this RQ, we consider the traditional self-attention mechanism and other recently proposed attention mechanisms~\cite{wang2020linformer}\cite{tay2020synthesizer}\cite{henry2020query}. Based on empirical results, we can select the most suitable mechanism for our proposed method {\tool}.





\subsection{Experimental Subjects}

In our empirical study, we choose two source code/pseudo code parallel corpora shared by Oda et al.~\cite{oda2015learning} (i.e., Django)  and Kulal et al.~\cite{kulal2019spoc} (i.e., SPoC). These corpora were gathered from real-world large-scale projects. Specifically, the code in Django is written by Python and the code in SPoC is written by C++.
To improve the quality of these two corpora, we remove the replicated code snippets to prevent that the same code snippet appears both in the training set and the test set. Then we divide the training set, validation set and test set by 80\%:10\%:10\%. 
Table~\ref{tab:statistics} and Table~\ref{tab:split} show the statistical information (such as average, mode, and median of the length, the percentage of code/pseudo-code when the length is smaller than 20, 50, and 100 respectively) in our used corpora.

\begin{table}[htbp] 
 \caption{Statistics of corpora used in our empirical study} 
 \label{tab:statistics}
 \vspace{-1mm}
 \begin{center}
 \begin{tabular}{ccccccc} 
  \toprule 
  \multicolumn{6}{c}{
  \textbf{Statistics for Code Length}} \\ 
    \midrule 
     & Avg & Mode & Median & $<20$ & $<50$ & $<100$ \\ 
    Django & 9.04 & 7 & 6 & $94.18\%$ & $99.66\%$ & $99.93\%$ \\ 
    SPoC & 9.51 & 6 & 8 & $93.14\%$ & $99.99\%$ & $100\%$ \\ 
     \midrule 
  \bottomrule 
  \multicolumn{6}{c}{
  \textbf{Statistics for Pseudo-code Length}} \\ 
    \midrule 
     & Avg & Mode & Median & $<20$ & $<50$ & $<100$ \\ 
    Django & 14.29 & 11 & 6 & $75.54\%$ & $93.12\%$ & $98.68\%$ \\ 
    SPoC & 8.53 & 4 & 7 & $95.66\%$ & $99.98\%$ & $100\%$ \\ 
  \bottomrule 
 \end{tabular}
 \end{center}
 \vspace{-1mm}
\end{table}

\begin{table}[htbp] 
 \caption{Corpora split results in our empirical study}
 \vspace{-1mm}
 \label{tab:split}
 \begin{center}
 \begin{tabular}{cccc} 
  \toprule 
  \textbf{Corpus} & \textbf{Training}  & \textbf{Validation}  & \textbf{Test}   \\
  \midrule 
Django  & 11,826 & 1,590  & 1,590  \\
SPoC     & 98,673 & 12,334 & 12,334 \\
  \bottomrule 
 \end{tabular} 
 \vspace{-1mm}
 \end{center}
\end{table}

\subsection{Performance Measures}

To quantitatively compare the performance between our proposed method and the baselines, we choose the following four performance measures, which have been widely used in previous neural machine translation studies~\cite{yang2020survey}. Notice the higher the value of the performance measure, the better the performance of the corresponding method. 
To ensure the implementation correctness of these performance measures, we use the nlg-eval library\footnote{\url{https://github.com/Maluuba/nlg-eval}}, which implements these performance measures for  natural language generation tasks.

\noindent\textbf{BLEU.} BLEU (Bilingual Evaluation Understudy)~\cite{papineni2002bleu} is a variant of the precision measure. This performance measure can calculate the similarity by computing the $n$-gram precision of a candidate sentence to the
reference sentence, with a penalty for the overly short length.


 
\noindent\textbf{METEOR.} METEOR (Metric for Evaluation of sentences with Explicit Ordering)~\cite{banerjee2005meteor} uses knowledge sources (such as WordNet) to expand the synset while taking into account the shape of words.


\noindent\textbf{ROUGE.}
ROUGE (Recall-Oriented Understudy for Gisting Evaluation)~\cite{lin2004rouge} is a set of indicators for evaluating automatic abstracts and machine sentences. This study uses ROUGE-L as our evaluation measure, where $L$ means LCS (longest common subsequence). 

\noindent\textbf{CIDER.} CIDER (Consensus-based Image Description Evaluation)~\cite{vedantam2015cider} is a collection of BLEU and vector space model. It treats each sentence as a document, and then calculates the tf-idf value of the $n$-gram phrase, and calculates the similarity between the candidate text and the reference text through the cosine distance. Finally, the average value is calculated based on $n$-grams of different lengths and used as the final result.


\subsection{Implementation Details and Running Platform}

In our empirical study, we use Pytorch 1.6.0 to implement our proposed method.
Our proposed method's hyper-parameters are mainly in three categories (i.e., the  hyper-parameters for the model structure, the hyper-parameters in the model training phase, and the hyper-parameters in the model test phase). The hyper-parameters for each category and their value are summarized in Table~\ref{Hyper-parameters}, where n\_layers means the number of layers (i.e., the depth of the model), d\_model means the model size (i.e., the width of the model).

All the experiments run on a computer with an Inter(R) Core(TM) i7-9750H 4210 CPU and a GeForce GTX1660ti GPU with 6 GB memory. The running OS platform is Windows 10.


\begin{table}[htbp]
 \caption{Hyper-parameters and their value in our empirical study}
 \vspace{-1mm}
 \begin{center}
\begin{tabular}{ccc}
\toprule
        \textbf{Category}               & \textbf{Hyper-parameter} & \textbf{Parameter Value} \\ \midrule
\multirow{6}{*}{Model Structure} & n\_layers       & 3     \\
                       & n\_heads        & 8     \\
                       & d\_model        & 256   \\
                       & hidden\_size    & 512   \\
                       & kernel\_size    & 5     \\
                       & scale           & $\sqrt{0.5}$   \\ \midrule
\multirow{4}{*}{Model Training Phase} & dropout         & 0.25  \\
                       & optimizer       & Adam  \\
                       & learning rate   & 0.001 \\ 
                       & batch size      & 128    \\ \midrule
\multirow{1}{*}{Model Test Phase}  & beam size       & 3    \\
  \bottomrule
\end{tabular}
 \end{center}
 \vspace{-1mm}
 \label{Hyper-parameters}
\end{table}


%% file: 5result.tex
\section{Result Analysis}
\label{sec:results}

\subsection{Result Analysis for RQ1}

\noindent\textbf{RQ1: Can our proposed method {\tool} outperform state-of-the-art baselines?}

\noindent\textbf{Method.}
In this RQ, we first compare our proposed method {\tool} with three state-of-the-art methods (i.e., Reduced-T2SMT method~\cite{oda2015learning}, Code-NN method~\cite{iyer2016summarizing} and Code2NL method~\cite{deng2020code}) in pseudo-code generation domain. 
Since these three studies all shared their code, we re-run their code on our corpora for a fair comparison. Then we also choose four baselines from the neural machine translation domain. 
According to the previous studies' description, we re-implemented these four baselines. Moreover, the value of hyper-parameters is consistent with that of our proposed method {\tool} for a fair comparison. Note that Reduced-T2SMT and Code2NL are customized with syntactic templates for python syntax, therefore these two methods cannot be applied to SPoC corpus, since the code is written by C++.
 The details of these baselines are summarized as follows.

(1) Reduced-T2SMT. This baseline~\cite{oda2015learning} is based on the framework of statistical
machine translation, which can generate natural language expressions for Python source codes.

(2) Code-NN. This baseline~\cite{iyer2016summarizing} is the first deep learning-based method, which uses LSTM with the attention mechanism to generate the summary of code snippets.

(3) Code2NL. This baseline~\cite{deng2020code} is an asynchronous learning model, which learns the code semantics and generates a fine-grained natural language description for each line of code. It adopts a type-aware sketch-based sequence-to-sequence learning method.

(4) RNN-seq2seq with (w)/without (w/o) attention mechanism. These RNN-based baselines~\cite{sutskever2014sequence}\cite{bahdanau2014neural} 
use RNN
to construct seq2seq models, and we compare the performance of these models with or without the attention mechanism. 

(5) ConvS2S. This baseline~\cite{gehring2017convolutional} uses CNN to construct seq2seq models. We also compare the performance of these models with or without the attention mechanism.

(6) Transformer. This baseline~\cite{vaswani2017attention} only uses encoder-decoder framework and the attention mechanism to achieve good performance. The most significant advantage of this baseline is that it can be efficiently parallelized.

\noindent\textbf{Results.} 
The comparison results between our proposed method {\tool} and baselines can be found in Table~\ref{tab:RQ1}. 
For Django,  
in terms of BLEU, {\tool} can improve the performance by 9.19\% to 39.29\%. 
In terms of METEOR, {\tool} can improve the performance by 0.84\% to 32.04\%. 
In terms of ROUGE-L, {\tool} can improve the performance by 0.27\% to 16.64\%.
In terms of CIDER, {\tool} can improve the performance by 14.90\% to 51.82\%.
For SPoC, 
in terms of BLEU, {\tool} can improve the performance by 6.21\% to 44.69\%. 
In terms of METEOR, {\tool} can improve the performance by 4.31\% to 39.70\%. 
In terms of ROUGE-L, {\tool} can improve the performance by 2.48\% to 13.83\%.
In terms of CIDER, {\tool} can improve the performance by 6.11\% to 35.79\%.

\begin{table}[htbp]
 \caption{The comparison results between our proposed method \tool and baseline methods}
 \vspace{-1mm}
 \begin{center}
 \setlength{\tabcolsep}{1mm}{
  \resizebox{0.5\textwidth}{!}{
 \begin{tabular}{cccccc}
  \toprule
\textbf{DATASET} & \textbf{METHOD} & \textbf{BLEU-4(\%)} & \textbf{METEOR(\%)} & \textbf{ROUGE-L(\%)} & \textbf{CIDER}  \\
  \midrule
\multirow{8}{*}{Django} &  Reduced-T2SMT  & 45.082  & 37.072 & 77.185 & 5.021 \\
 &  Code-NN  & 40.512  & 30.705 & 70.471 & 4.209 \\
 &  Code2NL  & 46.541  & 37.460 & 77.562 & 5.186 \\
 &  Seq2Seq w/o Atten. & 36.483  & 31.431 & 67.141 & 3.710 \\
 &  Seq2Seq w Atten.  & 43.960  & 38.874 & 73.036 & 4.902 \\
 &  ConvS2S  & 37.455  & 29.689 & 66.679 & 3.829 \\
 &  Transformer  & 43.464  & 35.466 & 72.910 & 5.059 \\
& {\tool} & \textbf{50.817} & \textbf{39.201} & \textbf{77.773} & \textbf{5.813}  \\
\midrule
\multirow{6}{*}{SPoC} 
 &  Code-NN  & 32.105  & 29.211 & 55.721 & 3.029 \\
 &  Seq2Seq w/o Atten. & 33.761  & 34.037 & 56.611 & 3.078 \\
 &  Seq2Seq w Atten.  & 41.007  & 37.892 & 59.852 & 3.556 \\
 &  ConvS2S  & 34.197  & 33.323 & 56.546 & 3.150 \\
 &  Transformer  & 43.738  & 39.124 & 61.896 & 3.876 \\
& {\tool} & \textbf{46.454} & \textbf{40.809} & \textbf{63.430} & \textbf{4.113}  \\
  \bottomrule
 \end{tabular} }
 }
 \end{center}
 \vspace{-1mm}
 \label{tab:RQ1}
\end{table}

\begin{tcolorbox}[width=1.0\linewidth, title={}]
\textbf{Summary for RQ1:}
Our proposed {\tool} can outperform baselines in terms of four performance metrics both in Django and SPoC.
\end{tcolorbox}
\vspace{-1mm}

\subsection{Result Analysis for RQ2}

\noindent\textbf{RQ2: Whether using the code feature extractor can improve the performance of {\tool}?}

\noindent\textbf{Method.} To verify whether using code feature extractor can improve the performance of {\tool}, we compared the performance between {\tool} with CFER and {\tool} without CFER.
Here, for these methods, the values of the hyper-parameters are consistent and all use the norm-attention as the attention mechanism (detailed discussion on the influence of different attention mechanisms on {\tool} can be found in Section~\ref{sec:resultRQ3}).

\noindent\textbf{Results.} The comparison results between {\tool} with (w) CFE and {\tool} without (w/o)  CFER can be found in Table~\ref{tab:new_RQ2}. 
For Django,
in terms of BLEU, {\tool} using CFER can improve the performance by 18.83\%.
In terms of METEOR, {\tool} using CFER can improve the performance by 1.16\%. 
In terms of ROUGE-L, {\tool} using CFER can improve the performance by 5.83\%.
In terms of CIDER, {\tool} using CFER can improve the performance by 14.84\%.
For SPoC,
in terms of BLEU, {\tool} using CFER can improve the performance by 5.18\%.
In terms of METEOR, {\tool} using CFER can improve the performance by 0.32\%. 
In terms of ROUGE-L, {\tool} using CFER can improve the performance by 1.01\%.
In terms of CIDER, {\tool} using CFER can improve the performance by 4.36\%.

\begin{table}[htbp]
 \caption{The comparison results between {\tool} with CFER  and {\tool} without CFER}
 \vspace{-1mm}
 \begin{center}
 \setlength{\tabcolsep}{1mm}{
 \resizebox{0.5\textwidth}{!}{
 \begin{tabular}{cccccc}
  \toprule
\textbf{DATASET} & \textbf{Name} & \textbf{BLEU(\%)} & \textbf{METEOR(\%)} & \textbf{ROUGE(\%)} & \textbf{CIDER}  \\
  \midrule
\multirow{2}{*}{Django} &   w/o CFER  & 42.765  & 38.752 & 73.492 & 5.062 \\
&  w CFER  & \textbf{50.817} & \textbf{39.201} & \textbf{77.773} & \textbf{5.813}  \\
  \midrule
 \multirow{2}{*}{SPoC} &  w/o CFER  & 44.168  & 40.677 & 62.799 & 3.941 \\
&  w CFER  & \textbf{46.454} & \textbf{40.809} & \textbf{63.430} & \textbf{4.113}  \\
  \bottomrule
 \end{tabular} }}
 \end{center}
 \vspace{-1mm}
 \label{tab:new_RQ2}
\end{table}

\begin{tcolorbox}[width=1.0\linewidth, title={}]
\textbf{Summary for RQ2:}
Using the code feature extractor in our proposed method {\tool} can improve the performance by effectively capturing code features.
\end{tcolorbox}
\vspace{-1mm}

\subsection{Result Analysis for RQ3}
\label{sec:resultRQ3}

\noindent\textbf{RQ3: What is the effect of attention mechanism on our proposed method {\tool}?}

\noindent\textbf{Method.}
In this RQ, we consider the impact of four different attention mechanisms on pseudo-code generation task. 
In particular, we consider the traditional attention mechanism (i.e., the self-attention mechanism) and state-of-the-art attention mechanisms (i.e., the other three attention mechanisms~\cite{wang2020linformer}\cite{tay2020synthesizer}\cite{henry2020query}).
We summarize the characteristics of the considered attention mechanisms as follows.

(1) Self-Attention.
Self-Attention~\cite{vaswani2017attention} essentially introduces contextual information for the current word to enhance the representation of the current word and write more information.

\begin{equation}
\text{ Self-Attention }(Q, K, V)=\operatorname{softmax}\left(\frac{Q K^{T}}{\sqrt{d_{k}}}\right) V
\end{equation}
where $Q$ is query, $K$ is key, $V$ is value, $Q, K, V \in R^{n\times d_{m}}$, they are the input word embedding matrix, $n$ is the length of the sentence, and $d_{m}$ is the dimension of the word embedding. The result of the dot product of $Q$ and $K$ is used to reflect the degree of influence of the context word on the central word, and then normalized by softmax.

(2) Linear-Attention.
By using linear multi-head attention~\cite{wang2020linformer}, the self-attention mechanism can be approximated by a low-rank matrix. Therefore, this kind of attention mechanism can reduce the overall self-attention time complexity and space complexity from $O(n^{2})$ to $O(n)$.

\begin{equation}
\text { Linear-Attention }(Q, K, V)=\operatorname{softmax}\left(\frac{Q (EK)^{T}}{\sqrt{d_{k}}}\right) FV
\end{equation}

For the Linear-Attention, two linear projection matrices $E, F \in R^{n \times k}$ are introduced when calculating key and value. This reduces the time and space complexity of the algorithm from $O(n^{2})$ to $O(nk)$. If we can choose a small projection dimension $k<<n$, we can significantly reduce the time and space complexity.
 
(3) Synthesizer-Attention.
Synthesizer-Attention~\cite{tay2020synthesizer} no longer calculates the dot product between two tokens vectors but learns to synthesize a self-alignment matrix (i.e., a synthetic self-attention matrix). Synthesizer is a generalization of the standard Transformer. The input of the model is $X\in R^{l \times d}$, and the output is $Y\in R^{l \times d}$, where $l$ is the length of the sentence and $d$ is the dimension of word embedding. The whole process can be divided into two steps: In the first step, it calculates the attention weight of each token and uses the parameterized function $F(X)$ to map the dimension of $X$ from the $d$ dimension to the $l$ dimension.
\begin{equation}
\text {B}=\text { F(X) }=W_{2}\left(\sigma_{R}\left(W_{1}(X)+b\right)\right)+b
\end{equation}
In the second step, it calculates the output result based on the attention score.
\begin{equation}
Y=\operatorname{Softmax}(B) G(X)
\end{equation}
where $B\in R^{l \times l}$,$G(X)$ is another parameterized function, which is analogous to $V$ in the self-attention mechanism.

(4) Norm-Attention.
Norm-Attention~\cite{henry2020query} applies L2 normalization to $Q$ and $K$, therefore $Q$ and $K$ become $\hat{Q}$ and $\hat{K}$:
\begin{equation}
\hat{Q}_{i}=\frac{Q_{i}}{\left\|Q_{i}\right\|},
\hat{K}_{i}=\frac{K_{i}}{\left\|K_{i}\right\|}
\end{equation}

Different from dividing by $\sqrt{d_k}$ in the Self-Attention mechanism, the Norm-Attention mechanism uses a learnable parameter to expand. The initial value of this parameter depends on the length of the sequence in the training data.
\begin{equation}
g_{0}=\log _{2}\left(L^{2}-L\right)
\end{equation}
where $L$ is the 97.5th percentile sequence length of all training data in the source and the target, therefore the formula of Norm-Attention is defined as follows.

\begin{equation}
\text { Norm-Attention }(Q, K, V)=\operatorname{softmax}\left(g * \hat{Q} \hat{K}^{T}\right) V
\end{equation}

\begin{figure*}[htbp]
	\centering
    \vspace{-1mm}
	\includegraphics[width=0.9\textwidth]{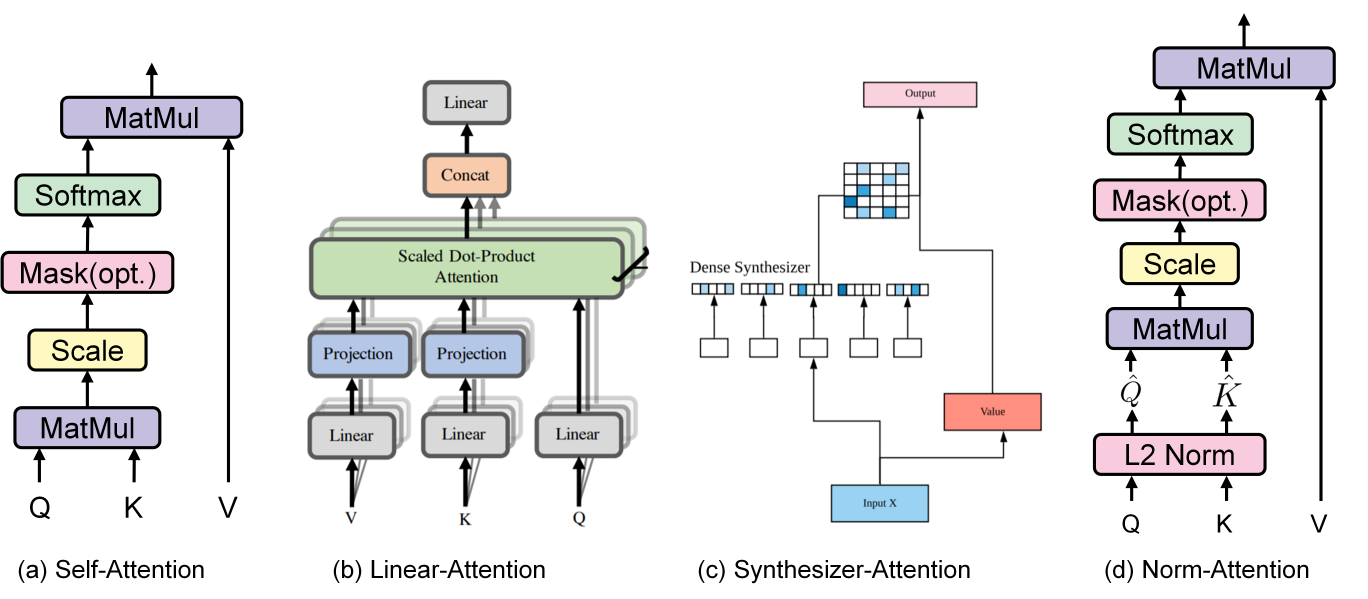}
	\caption{Structure of different attention mechanisms}
    \vspace{-1mm}
	\label{fig:attentionstructure}
\end{figure*}

The structure of these four different attention mechanisms can be found in Fig.~\ref{fig:attentionstructure}.

\noindent\textbf{Results.} 
The comparison results between different attention mechanisms in our proposed method {\tool} can be found in Table~\ref{tab:RQ2}. In terms of a specific performance measure, we mark the best result in the bold type.
For Django,
in terms of BLEU, the Norm-Attention mechanism can improve the performance by 1.40\% to 6.16\%. 
In terms of METEOR, the Norm-Attention mechanism can improve the performance by 2.34\% to 6.92\%. 
In terms of ROUGE-L, the Norm-Attention mechanism can improve the performance by 0.41\% to 5.84\%.
In terms of CIDER, the Norm-Attention mechanism can improve the performance by 3.60\% to 26.29\%.
For SPoC,
in terms of BLEU, the Norm-Attention mechanism can improve the performance by 0.74\% to 5.06\%. 
In terms of METEOR, the Norm-Attention mechanism can improve the performance by 3.83\% to 8.71\%. 
In terms of ROUGE-L, the Norm-Attention mechanism can improve the performance by 1.56\% to 4.88\%.
In terms of CIDER, the Norm-Attention mechanism can improve the performance by 5.16\% to 14.15\%.

Moreover, we surprisingly find that the two state-of-the-art attention mechanisms (i.e., the Linear-Attention mechanism and the Synthesizer-Attention mechanism) cannot achieve better performance than the original Self-Attention mechanism. Therefore, we recommend the usage of the Norm-Attention mechanism in our proposed method {\tool}. Therefore,  except for low-resource machine translation tasks~\cite{henry2020query}, the Norm-Attention mechanism is also applicable to our studied pseudo-code generation problem.

\begin{table}[htbp]
 \caption{The comparison results between different attention mechanisms}
 \begin{center}
 \setlength{\tabcolsep}{1mm}{
 \resizebox{0.5\textwidth}{!}{
 \begin{tabular}{cccccc}
  \toprule
\textbf{DATASET} & \textbf{Name} & \textbf{BLEU(\%)} & \textbf{METEOR(\%)} & \textbf{ROUGE(\%)} & \textbf{CIDER}  \\
  \midrule
\multirow{4}{*}{Django} &   Self-Attention  & 50.115  & 38.305 & 77.458 & 5.611 \\
  & Linear-Attention  & 48.215  & 37.541 & 75.252 & 5.077 \\
  & Synthesizer-Attention  & 47.868  & 36.664 & 73.481 & 4.603 \\
  & Norm-Attention  & \textbf{50.817} & \textbf{39.201} & \textbf{77.773} & \textbf{5.813} \\
  \midrule
  \multirow{4}{*}{SPoC} &   Self-Attention  & 46.115  & 39.305 & 62.458 & 3.911 \\
  & Linear-Attention  & 44.215  & 37.541 & 61.252 & 3.717 \\
  & Synthesizer-Attention  & 45.868  & 37.664 & 60.481 & 3.603 \\
  & Norm-Attention  & \textbf{46.454} & \textbf{40.809} & \textbf{63.430} & \textbf{4.113} \\
  \bottomrule
 \end{tabular}} }
 \end{center}
 \label{tab:RQ2}
\end{table}

\begin{tcolorbox}[width=1.0\linewidth, title={}]
\textbf{Summary for RQ3:}
Using the Norm-Attention mechanism can achieve the best performance in our proposed {\tool}.
\end{tcolorbox}
\vspace{-1mm}

%% file: 6discuss.tex
\section{Discussions}
\label{sec:discuss}


In this section, we further conduct experiments by analyzing the parameter effects on {\tool}. These parameters are the model size $d\_model$, the number of layers $N$, and the number of kernel\_size $K$. The comparison results between different hyperparameters can be found in Table~\ref{tab:dis}. 
The optimal value of these parameters is set as follows. $d\_model$ is 256, $N$ is 3, and $K$ is 5.
When analyzing the impact of a specific hyperparameter, we keep other hyperparameters' value as the optimal value.

\begin{table}[htbp]
  \begin{center}
 \caption{Sensitivity analysis on the hyperparameters (i.e., the model size, the number of layers, and the number of kernel\_size)}
 \label{tab:dis}

 \resizebox{0.5\textwidth}{!}{
\begin{tabular}{cccccc}
\toprule
\multicolumn{1}{c|}{DATASET}                 & \multicolumn{1}{c|}{VALUE} & BLEU(\%) & METEOR(\%) & ROUGE-L(\%) & CIDER \\ \midrule
\multicolumn{6}{c}{Varying the model size($d\_model$)}                                                        \\ \midrule
\multicolumn{1}{c|}{\multirow{3}{*}{Django}} & \multicolumn{1}{c|}{256}   &     \textbf{50.817} & \textbf{39.201} & \textbf{77.773} & \textbf{5.813} \\
\multicolumn{1}{c|}{}                        & \multicolumn{1}{c|}{512}   &      47.310&   37.349 &  75.208 & 5.396  \\
\multicolumn{1}{c|}{}                        & \multicolumn{1}{c|}{768}   &    0  &  2.279    &   5.066    &   0.004 \\ \midrule
\multicolumn{1}{c|}{\multirow{3}{*}{SPoC}}   & \multicolumn{1}{c|}{256}   &     \textbf{46.454} & \textbf{40.809} & \textbf{63.430} & \textbf{4.113}  \\
\multicolumn{1}{c|}{}                        & \multicolumn{1}{c|}{512}   &     36.185 &   38.241    &  61.742   &   3.764  \\
\multicolumn{1}{c|}{}                        & \multicolumn{1}{c|}{768}   &     0 &  1.242 &   0.624    &   0.003 \\ \midrule
\multicolumn{6}{c}{Varying the number of layers($N$)}                                                         \\ \midrule
\multicolumn{1}{c|}{\multirow{3}{*}{Django}} & \multicolumn{1}{c|}{3}     &      \textbf{50.817} & \textbf{39.201} & \textbf{77.773} & \textbf{5.813} \\
\multicolumn{1}{c|}{}                        & \multicolumn{1}{c|}{6}     &      44.815  &  35.671    &   73.409  &  4.944\\
\multicolumn{1}{c|}{}                        & \multicolumn{1}{c|}{9}     &    0  &  2.279    &   5.066    &   0.004   \\ \midrule
\multicolumn{1}{c|}{\multirow{3}{*}{SPoC}}   & \multicolumn{1}{c|}{3}     &     \textbf{46.454} & \textbf{40.809} & \textbf{63.430} & \textbf{4.113}  \\
\multicolumn{1}{c|}{}                        & \multicolumn{1}{c|}{6}     &     33.860 &  35.935  &   57.176  &  3.327  \\
\multicolumn{1}{c|}{}                        & \multicolumn{1}{c|}{9}     &     0 &  1.242 &   0.624    &   0.003     \\ \midrule
\multicolumn{6}{c}{Varying the number of kernel\_size($K$)}                                                   \\ \midrule
\multicolumn{1}{c|}{\multirow{3}{*}{Django}} & \multicolumn{1}{c|}{3}     &     49.799 &  38.706   &   76.976   &   5.739 \\
\multicolumn{1}{c|}{}                        & \multicolumn{1}{c|}{5}     &     \textbf{50.817} & \textbf{39.201} & 77.773 & \textbf{5.813} \\
\multicolumn{1}{c|}{}                        & \multicolumn{1}{c|}{7}     &      50.551&     39.165   &  \textbf{77.795}   &   5.799 \\ \midrule
\multicolumn{1}{c|}{\multirow{3}{*}{SPoC}}   & \multicolumn{1}{c|}{3}     &      39.706&  39.690   &  59.852   &  3.731  \\
\multicolumn{1}{c|}{}                        & \multicolumn{1}{c|}{5}     &     \textbf{46.454} & \textbf{40.809} & \textbf{63.430} & \textbf{4.113}  \\
\multicolumn{1}{c|}{}                        & \multicolumn{1}{c|}{7}     &      35.661 &  36.629 &   57.444  &    3.352 \\   \bottomrule
\end{tabular}}
 \end{center}
\end{table}

The final results can be found in Table~\ref{tab:dis}. We find that it is harder to train the generation model when $N$ is large (i.e., depth of the model) and $d\_model$ is large (i.e., width of the model). This is because the gradient disappearance problem tends to occur when training the generation model, even though the {\tool} adds layer normalization, gating mechanism, and residual connection to alleviate this problem. This is the reason why we choose small value for these network parameters, which are more suitable for this pseudo-code generation task. Regarding the value of $kernel\_size$, We find that the performance of the model will decrease if the value of this parameter is large or small. Therefore, we set the value of the parameter $kernel\_size$ to 5 in {\tool}.

%% file: 7threats.tex
\section{Threats to validity}
\label{sec:threat}

In this section, we mainly analyze the potential threats to the validity of our empirical study.

\noindent\textbf{Internal Threats.} 
The first internal threat is the potential defects in the implementation of {\tool}. 
To alleviate this threat, we check our implementation carefully and use mature libraries, such as PyTorch and TorchText.
The second internal threat is the implementation of chosen baselines. 
To alleviate this threat, we use the implementation shared by previous pseudo-code generation studies~\cite{oda2015learning}.

\noindent\textbf{External Threats.} 
The main external threat is the choice of corpora. To alleviate this threat, we select the popular two corpora provided by Oda et al.~\cite{oda2015learning} and Kulal et al.~\cite{kulal2019spoc} and these corpora have been widely used in previous studies on pseudo-code generation. In the future, we want to gather more corpora from other commercial or open-source projects and verify the effectiveness of our proposed method {\tool}.

\noindent\textbf{Construct Threats.} 
The construct threat in this study is the performance measures used to evaluate the performance of {\tool}.
To alleviate these threats, we choose three performance measures, which have been used by the previous pseudo-code generation studies. Moreover, we also consider a performance measure, CIDER, which has been more used in recent code comment generation studies~\cite{hu2020deep}\cite{li2021secnn}\cite{yang2021comformer}.

\noindent\textbf{Conclusion Threats.} 
The conclusion threat in our study is we do not perform cross-validation (CV) in our research.
Using CV can comprehensively evaluate our proposed method {\tool}, since different split may result in a diverse training set, validation set, and testing set. However, this model evaluation method has not been commonly used for neural machine translation due to high training computational cost. 

%% file: 8conclusion.tex
\section{Conclusion and Future Work}
\label{sec:conclusion}

In this study, we propose a novel deep pseudo-code generation method {\tool} via Transformer and code feature extraction. 
In the encoder part, {\tool} utilizes both Transformer encoder and code feature extractor to perform encoding for source code. In the decoder part, {\tool} resorts to the beam search algorithm for pseudo-code generation.
Empirical results verified the effectiveness of our proposed method.
Moreover, we also design the experiments to justify the component setting's rationality in {\tool}.

In the future, we want to improve the effectiveness of our proposed method {\tool} by considering more advanced attention mechanisms and pre-training models. Moreover, we also want to verify the effectiveness of {\tool} by considering corpora gathered from other commercial or open-source projects developed by other programming languages.